\documentclass{INTERSPEECH2023}

\interspeechcameraready

\usepackage{caption}
\usepackage{subcaption}
\usepackage{hyperref}
\usepackage{graphicx}

\title{VITS2: Improving Quality and Efficiency of Single-Stage Text-to-Speech with Adversarial Learning and Architecture Design}
\name{Jungil Kong, Jihoon Park, Beomjeong Kim,  Jeongmin Kim, Dohee Kong, Sangjin Kim}

\address{
  SK Telecom, South Korea}
\email{\texttt{\{jik876,batho2n,beomjeong.kim,jmkim94,dohee.kong,kimsangjin\}@sk.com}}

\begin{document}

\maketitle

\begin{abstract}
Single-stage text-to-speech models have been actively studied recently, and their results have outperformed two-stage pipeline systems. Although the previous single-stage model has made great progress, there is room for improvement in terms of its intermittent unnaturalness, computational efficiency, and strong dependence on phoneme conversion. In this work, we introduce VITS2, a single-stage text-to-speech model that efficiently synthesizes a more natural speech by improving several aspects of the previous work. We propose improved structures and training mechanisms and present that the proposed methods are effective in improving naturalness, similarity of speech characteristics in a multi-speaker model, and efficiency of training and inference. Furthermore, we demonstrate that the strong dependence on phoneme conversion in previous works can be significantly reduced with our method, which allows a fully end-to-end single-stage approach.
\end{abstract}

\section{Introduction}
\label{sec:Introduction}
Recent developments in deep neural network-based text-to-speech have seen significant advancements. Deep neural network-based text-to-speech is a method for generating corresponding raw waveforms from input texts; it has several interesting features that often make the text-to-speech task challenging. A quick review of the features reveals that the text-to-speech task involves converting text, which is a discontinuous feature, into continuous waveforms. The input and output have a time step difference of hundreds of times, and the alignment between them must be very precise to synthesize high-quality speech audio. Additionally, prosody and speaker characteristics not present in the input text should be expressed naturally and it is a one-to-many problem in which text input can be spoken in multiple ways. Another factor that makes synthesizing high-quality speech challenging is that humans focus on individual components when listening to an audio; therefore, even if a fraction of the hundreds of thousands of signals that constitute the entire audio are unnatural, humans can easily sense them. Efficiency is another factor that makes the task difficult. The synthesized audio has a substantial time resolution, which generally comprises more than 20,000 data per second, demanding highly efficient sampling methods.

Owing to the text-to-speech task features, the solution can also be sophisticated. Previous works have addressed these problems by dividing the process of generating waveforms from input texts into two cascaded stages. A popular method involves producing intermediate speech representations such as mel-spectrograms or linguistic features from the input texts in the first stage \cite{shen2018natural,li2019neural,ren2019fastspeech,kim2020glow,valle2020flowtron,popov2021grad,ren2021fastspeech} and then generating raw waveforms conditioned on those intermediate representations in the second stage \cite{oord2016wavenet,kalchbrenner2018efficient,prenger2019waveglow,kumar2019melgan,binkowski2019high,yamamoto2020parallel,kong2020hifi,chen2020wavegrad}.
Two-stage pipeline systems have the advantages of simplifying each model and facilitating training; however, they also have the following limitations.
1) Error propagation from the first stage to the second stage.
2) Rather than utilizing the learned representation inside the model, it is mediated through human-defined features such as mel-spectrogram or linguistic features.
3) Computation required to generate intermediate features.
Recently, to address these limitations, single-stage models that directly generate waveforms from input texts have been actively studied \cite{donahue2021endtoend, ren2021fastspeech, kim2021conditional, lim22_interspeech}.
The single-stage models not only outperformed the two-stage pipeline systems, but also showed an ability to generate high-quality speech nearly indistinguishable from humans.

Although the previous work \cite{kim2021conditional} has achieved great success with the single-stage approach, the model \cite{kim2021conditional} has the following problems: intermittent unnaturalness, low efficiency of the duration predictor, complex input format to alleviate the limitations of alignment and duration modeling (use of blank token), insufficient speaker similarity in the multi-speaker model, slow training, and strong dependence on the phoneme conversion.
In this work, we provide methods to address these problems.
We propose a stochastic duration predictor trained through adversarial learning, normalizing flows improved by utilizing the transformer block and a speaker-conditioned text encoder to model multiple speakers' characteristics better. We confirm that the proposed methods improve quality and efficiency. Furthermore, we show that the methods reduce the dependency on the phoneme conversion through the experiment using normalized texts as the input of the model. 
Thus, the methods move closer to a fully end-to-end single-stage approach.

\section{Method}
\label{sec:Method}
In this section, we describe improvements in four subsections: duration prediction, augmented variational autoencoder with normalizing flows, alignment search, and speaker-conditioned text encoder. We propose a method that uses adversarial learning to train the duration predictor to synthesize natural speech with high efficiency in both training and synthesis. Our model essentially learns alignments using the Monotonic Alignment Search (MAS) proposed in the previous work~\cite{kim2020glow, kim2021conditional}, and we further suggest a modification to improve the quality. In addition, we propose a method to improve naturalness by introducing the transformer block into the normalizing flows, which enables capturing long-term dependencies when transforming the distribution. Furthermore, we modify the speaker conditioning to improve the speaker similarity in a multi-speaker model.

\begin{figure*}[ht]
    \vskip -0.1in
    \begin{center}
        \hskip 0.0in
        \begin{subfigure}{.26\textwidth}
            \hspace{0.35cm}
            \includegraphics[width=.90\linewidth]{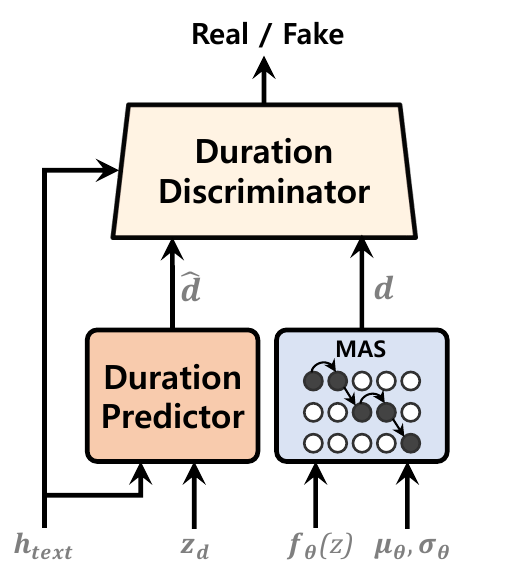}
            \vskip 0.0in
            \caption{Training of the duration \\ predictor}
            \label{fig:dp}
        \end{subfigure}
        \hskip 0.65in
        \begin{subfigure}{.18\textwidth}
            \centering
            \captionsetup{justification=centering}
            \includegraphics[width=.85\linewidth]{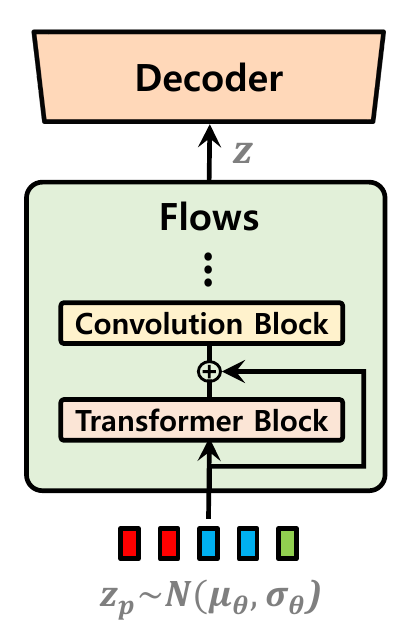}
            \vskip 0.0in
            \centering
            \caption{Normalzing flows with \\ the transformer block}
            \label{fig:nf}
        \end{subfigure}
        \hskip 0.45in
        \begin{subfigure}{.30\textwidth}
            \hspace{0.38cm}
            \includegraphics[width=.83\linewidth]{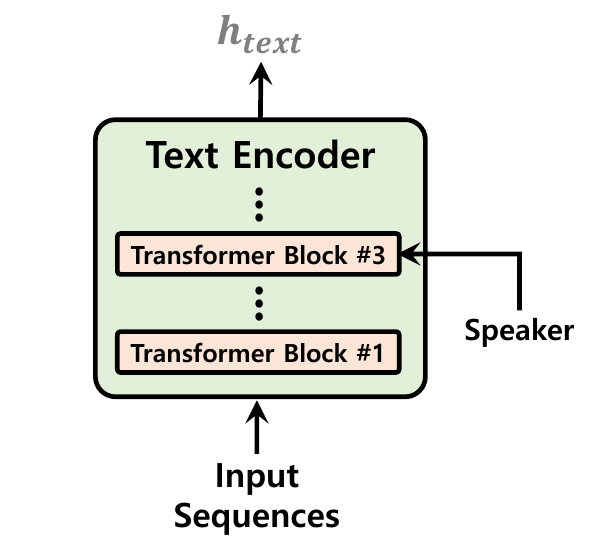}
            \vskip 0.015in
            \caption{Speaker-conditioned \quad\quad \ \\ text encoder \quad \ }
            \label{fig:scte}
        \end{subfigure}\hfill%
    \end{center}
    \vskip -0.15in
    \caption{Diagram depicting (a) Training of the duration predictor, (b) Normalizing flows with the transformer block, and (c) Speaker-conditioned text encoder.}
    \vskip -0.15in
\end{figure*}

\subsection{Stochastic Duration Predictor with Time Step-wise Conditional Discriminator}
\label{ssec:mehtod_sdpadv}
The previous work~\cite{kim2021conditional} has shown that the flow-based stochastic duration predictor is more effective in improving the naturalness of synthesized speech than the deterministic approach. It showed great results; however, the flow-based method requires relatively more computations and some sophisticated techniques. We propose a stochastic duration predictor with adversarial learning to synthesize more natural speech with higher efficiency in both training and synthesis than the previous work~\cite{kim2021conditional}. The overview of the proposed duration predictor and discriminator is shown in Figure~\ref{fig:dp}. We apply adversarial learning to train the duration predictor with a conditional discriminator that is fed the same input as the generator to appropriately discriminate the predicted duration. 
We use the hidden representation of the text $h_{text}$ and Gaussian noise $z_{d}$ as the input of the generator $G$; and the $h_{text}$ and duration obtained using MAS in the logarithmic scale denoted as $d$ or predicted from the duration predictor denoted as $\hat{d}$, are used as the input of the discriminator $D$.
Discriminators of general generative adversarial networks are fed inputs of a fixed length, whereas the duration for each input token is predicted, and the length of the input sequence varies for each training instance. To properly discriminate the inputs of variable length, we propose a time step-wise discriminator that discriminates each of the predicted durations of all tokens.
We use two types of losses; the least-squares loss function~\cite{mao2017least} for adversarial learning and the mean squared error loss function:
\begin{flalign}
    L_{adv}(D) &= \mathbb{E}_{(d,z_{d},h_{text})}\Big[(D(d,h_{text})-1)^2 \notag \\& 
    \quad\quad\quad\quad \ +(D(G(z_{d},h_{text}),h_{text}))^2\Big], \\
    L_{adv}(G) &= \mathbb{E}_{(z_{d},h_{text})}\Big[(D(G(z_{d},h_{text}))-1)^2\Big], \\
    \vspace{1cm} 
    L_{mse} &= MSE(G(z_{d},h_{text}), d) 
\end{flalign}
Our proposed duration predictor and training mechanism allow for a learning duration in short steps, and the duration predictor is separately trained as the last training step, which reduces the overall computation time for training.

\subsection{Monotonic Alignment Search with Gaussian Noise}
\label{ssec:mehtod_maswn}
Following the previous work~\cite{kim2020glow, kim2021conditional}, we introduce MAS into our model to learn the alignment. 
The algorithm yields the alignment between text and audio that has the highest probability among all possible monotonic alignments, and the model is trained to maximize its probability. The method is efficient; however, after searching and optimizing a particular alignment, it is limited in exploration to search for other alignments that are more appropriate.
To mitigate this, we add a small Gaussian noise to the calculated probabilities. This gives the model extra opportunities to search for other alignments. We only add this noise at the beginning of training because MAS enables the model to learn the alignments quickly. Referring to a previous work~\cite{kim2020glow}, which described the algorithm in detail, $Q$ values have the maximum log-likelihood calculated for all possible positions in the forward operation. We add small Gaussian noise $\epsilon$ to the calculated $Q$ values in the operation.
\begin{align}
P_{i,j} & = {\log \mathcal{N}(z_j; \mu_{i}, \sigma_{i}}) &&\\
Q_{i,j} & = \max_{A}{\sum_{k=1}^{j}{\log \mathcal{N}(z_k; \mu_{A(k)}, \sigma_{A(k)}}}) \notag &&\\
&= \max(Q_{i-1, j-1}, Q_{i, j-1}) + P_{i,j}+\epsilon 
\end{align}
where $i$ and $j$ denote a specific position on the input sequence and posterior, respectively, $z$ represents transformed latent variables from the normalizing flows. $\epsilon$ is obtained as the product of noise sampled from the standard normal distribution, the standard deviation of $P$, and the noise scale starting at 0.01 and decreasing by $2\times10^{-6}$ for every step.

\subsection{Normalizing Flows using Transformer Block}
\label{ssec:mehtod_nftr}
\begin{figure}[t]
    \begin{center}
            \centering
            \hspace{-0.5cm}
            \includegraphics[width=5cm]{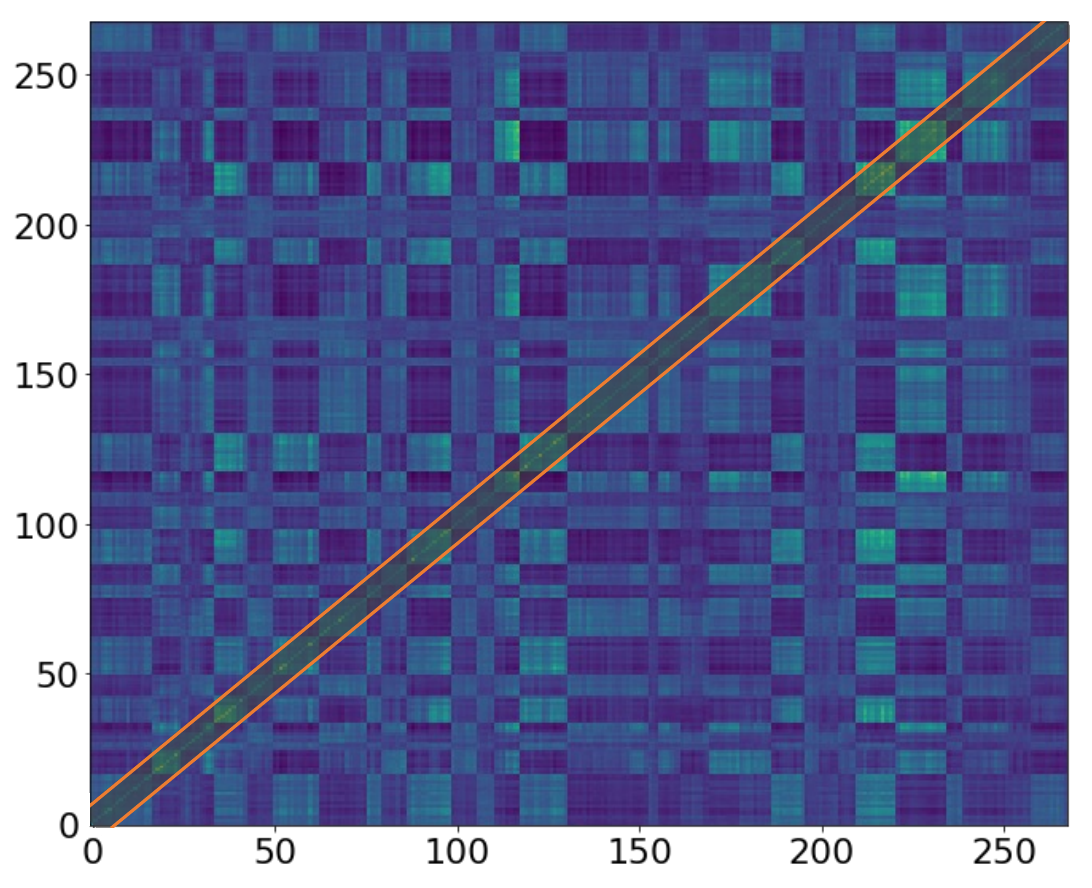}
    \end{center}
    \vspace{-0.4cm}
    \caption{Actual attention score map learned with the transformer block in the normalizing flows. The inside area of the orange lines corresponds to the receptive field of the first convolution block in the normalizing flows of the previous work~\cite{kim2021conditional}. It demonstrates that the transformer block collects information to transform the distribution at various positions that can not be captured with the convolution block.}
    \label{fig:nf_attn}
\vskip -0.18in
\end{figure}

The previous work~\cite{kim2021conditional} demonstrated the capability of the variational autoencoder augmented with normalizing flows to synthesize high-quality speech audio. The normalizing flows comprise convolution blocks, which are effective structures for capturing the patterns of adjacent data and enabling the model to synthesize high-quality speech. The ability to capture long-term dependencies can be crucial when transforming distribution because each part of the speech is related to other parts that are not adjacent. Although a convolution block captures adjacent patterns effectively, it has a disadvantage in capturing long-term dependencies owing to the limitations of its receptive field. Therefore, we add a small transformer block with the residual connection into the normalizing flows to enable the capturing of long-term dependencies, as shown in Figure~\ref{fig:nf}. Figure~\ref{fig:nf_attn} shows an actual attention score map and the receptive field of the convolution block. We can confirm that the transformer block collects information at various positions when transforming the distribution, which is impossible with the receptive field. 

\subsection{Speaker-Conditioned Text Encoder}
\label{ssec:mehtod_scte}
Because the multi-speaker model is to synthesize speech in multiple characteristics according to the speaker condition with one single model, expressing individual speech characteristics of each speaker is an important quality factor as well as naturalness. The previous work showed that the single-stage model can model multiple speakers with high quality. Considering some features, such as a speaker’s particular pronunciation and intonation, significantly influences the expression of the speech characteristics of each speaker but are not contained in the input text, we design a text encoder conditioned with the speaker information to better mimic various speech characteristics of each speaker by learning the features while encoding the input text. We condition the speaker vector on the third transformer block of the text encoder, as shown in Figure~\ref{fig:scte}.

\section{Experiments}
\label{sec:Experiments}
We conducted experiments on two different datasets. We used the LJ Speech dataset~\cite{ljspeech17} to confirm the improvement in naturalness and the VCTK dataset~\cite{veaux2017cstr} to verify whether our model could reproduce speaker characteristics better. 
The LJ Speech dataset consists of 13,100 short audio clips of a single speaker with a total length of approximately 24 hours. The audio format is 16-bit PCM with a sample rate of 22.05 kHz, and we used it without any manipulation. We randomly split the dataset into a training set (12,500 samples), validation set (100 samples), and test set (500 samples). The VCTK dataset consists of approximately 44,000 short audio clips uttered by 109 native English speakers with various accents. The total length of the audio clips is approximately 44 hours. The audio format is 16-bit PCM with a sample rate of 44.1 kHz. We reduced the sample rate to 22.05 kHz. We randomly split the dataset into a training set (43,470 samples), validation set (100 samples), and test set (500 samples).

We used 80 bands mel-scale spectrograms for calculating the reconstruction loss. In contrast with the previous work~\cite{kim2021conditional}, we used the same spectrograms as the input of the posterior encoder. The fast Fourier transform, window, and hop sizes were set to 1024, 1024, and 256, respectively.

We conducted experiments using both phoneme sequences and normalized texts as the input of the model. 
We converted text sequences into International Phonetic Alphabet sequences using open-source software~\cite{phonemizer20} and fed the text encoder with the sequences. Contrasting with the previous work~\cite{kim2021conditional}, we did not use the blank token. For the experiment with normalized texts, we normalized the input text with simple rules using open-source software~\cite{keithitotacotron} and fed the text encoder with it.

The networks were trained using the AdamW~\cite{loshchilov2018decoupled} optimizer with $\beta_1 = 0.8$, $\beta_2 = 0.99$, and weight decay $\lambda = 0.01$. The learning rate decay was scheduled by a $0.999^{1/8}$ factor in every epoch, with an initial learning rate of $2\times10^{-4}$. We fed the networks with 256 training instances per step.
Following the previous work~\cite{kim2021conditional}, the windowed generator training was applied. We used mixed precision training on four NVIDIA V100 GPUs. The networks to generate waveforms and the duration predictor were trained up to 800k and 30k steps, respectively.

\section{Results}
\label{sec:Results}
\subsection{Evaluation of Naturalness}
\label{ssec:results_mos}
To confirm that the proposed model synthesizes natural speech, crowdsourced mean opinion score (MOS) tests were conducted. Raters rated their naturalness on a 5-point scale from 1 to 5 after listening to randomly selected audio samples from the test sets. Considering that the previous work~\cite{kim2021conditional} has already demonstrated similar quality to human recordings, we also conducted a comparative mean opinion score (CMOS) test, which is appropriate for evaluating high-quality samples by direct comparison. Raters rated their relative preference in terms of naturalness on a 7-point scale from 3 to -3 after listening to randomly selected audio samples from the test sets.\footnote{Demo: \href{https://vits-2.github.io/demo/}{https://vits-2.github.io/demo/}}
Raters were allowed to evaluate each audio sample once. All audio samples were normalized to avoid the effect of amplitude differences on the score. We used the official implementation and pre-trained weights of the previous work~\cite{kim2021conditional} as the comparison model. The evaluation results are presented in Table~\ref{tab:mos} and Table~\ref{tab:cmos}. The MOS difference between our method and the previous work~\cite{kim2021conditional} was 0.09, and the CMOS and confidence interval were 0.201 and $\pm$0.105, respectively. The results demonstrate that the our method significantly improves the quality of synthesized speech. Additionally, we evaluated CMOS with the method~\cite{lim22_interspeech} that showed good performance using different structures and training mechanisms. For evaluation, we generated samples using the official implementation and pre-trained weights. The CMOS and confidence intervals of the evaluation are 0.176 and $\pm$0.125, respectively, indicating that our method significantly outperforms the method.

\subsection{Ablation Studies}
\label{ssec:ablation_studies}
Ablation studies were also conducted to verify the validity of the proposed methods. To verify the validity of the stochastic duration predictor trained with adversarial learning, it was substituted with the deterministic duration predictor that had the same structure and was trained with L2 loss. The deterministic duration predictor was trained up to the same steps as the previous work~\cite{kim2021conditional}. To verify the efficacy of the noise scheduling used in the alignment search, the model was trained without the noise. We trained the model without the transformer block in the normalizing flows to verify its effectiveness. The evaluation results are presented in Table~\ref{tab:mos}. The MOS differences of the ablation studies on the deterministic duration predictor, alignment search without the noise, and normalizing flows without the transformer block are 0.14, 0.15, and 0.06, respectively. As we do not use the blank token and linear spectrogram, the computational efficiency would be improved, and removing some of the proposed methods shows lower performance compared with the previous work~\cite{kim2021conditional}. The results show that the proposed methods are effective in improving the quality.

\begin{table}[t]
  \caption{Comparison of MOS of the proposed model, the previous work, and the ablation studies on the LJ Speech dataset with 95\% confidence intervals. }
  \label{tab:mos}
  \vspace{-0.8\baselineskip}
  \begin{center}
  \vskip -0.1in
  \begin{tabular}[htbp]{lc}
    \toprule
    Model   & 
    MOS (CI) \\
    \midrule
    Ground Truth    & 4.43 ($\pm$0.06) \\
    \midrule
    VITS    & 4.38 ($\pm$0.06) \\
    \midrule
    VITS2    & \textbf{4.47} ($\pm$0.06) \\
    \,\, w/o \scalebox{.86}[1.0]{Adversarial Learning for Duration Predictor} & 4.33 ($\pm$0.07) \\
    \,\, w/o Alignment Noise & 4.32 ($\pm$0.07) \\
    \,\, w/o Transformer Block & 4.41 ($\pm$0.07) \\
    \bottomrule
  \end{tabular}
  \end{center}
  \vspace{-0.5\baselineskip}
\end{table}

\begin{table}[t]
    \caption{(a) Comparative MOS of the proposed model and the previous works on the LJ Speech dataset with 95\% confidence intervals. (b) Comparison of similarity MOS of the proposed model and the previous work on the VCTK dataset with 95\% confidence intervals .}
    \centering
    \hskip -0.1in
    \vspace{-0.3\baselineskip}
    \subfloat[Comparative MOS]{%
      \label{tab:cmos}
      \vspace{-1.2\baselineskip}
      \begin{tabular}[htbp]{lcc}
        \toprule
        Model   & 
        CMOS   & 
        CI \\
        \midrule
        VITS & \textbf{0.201} & $\pm$0.105 \\
        JETS & \textbf{0.176} & $\pm$0.125 \\
        \bottomrule
      \end{tabular}
      }
  \quad
    \subfloat[Similarity MOS]{%
      \label{tab:smos}
      \begin{tabular}[htbp]{lc}
        \toprule
        Model   & 
        MOS (CI) \\
        \midrule
        VITS    & 3.79 ($\pm$0.09) \\
        VITS2    & \textbf{3.99} ($\pm$0.08) \\
        \bottomrule
      \end{tabular}
      }
      \vskip -0.01in
  \hspace{0.3cm}
\end{table}

\subsection{Evaluation of Speaker Similarity}
\label{ssec:results_smos}
To confirm the improvement in speaker similarity in the multi-speaker model, similarity MOS tests similar to the previous work~\cite{jia2018transfer} were conducted through crowdsourcing. In the test, randomly sampled human recorded audio from the test set was presented as a reference, and raters scored the similarity between the reference and the corresponding synthesized audio on a five-point scale from 1 to 5. As in section 4.1, raters were allowed to evaluate each audio sample once, and the audio samples were normalized.
The evaluation results are presented in Table~\ref{tab:smos}. VITS2 was rated 0.2 MOS higher than the previous work~\cite{kim2021conditional}, which shows the effectiveness of our method in improving speaker similarity when modeling multiple speakers.

\subsection{Reduced dependency on the phoneme conversion}
\label{ssec:results_phoneme}
Previous works~\cite{kim2021conditional,tan2022naturalspeech} have shown good performance with single-stage approaches but continue to have a strong dependence on phoneme conversion. Because normalized text does not inform its actual pronunciation, it makes learning accurate pronunciations challenging. It is currently a crucial barrier to achieving a fully end-to-end single-stage speech synthesis. We present that our method significantly improves this problem through intelligibility tests. After transcribing 500 synthesized audio in the test set using Google's automatic speech recognition API, we calculated the character error rate (CER) with the ground truth text as the reference. We compared the results of the following four models with the ground truth: the proposed model using phoneme sequences, the proposed model using normalized texts, the previous work using phoneme sequences, and the previous work using normalized texts. Table~\ref{tab:intelligibility_test} presents the comparison,
which confirms that not only the proposed model outperforms the previous work, but also the performance of our model using normalized texts is comparable to that of the model using phoneme sequences. It demonstrates the possibility of a data-driven, fully end-to-end approach.

\begin{table}[t]
  \caption{Comparison of the intelligibility tests on the LJ Speech dataset.}
  \label{tab:intelligibility_test}
  \vspace{-0.5\baselineskip}
  \begin{center}
  \vskip -0.1in
  \begin{tabular}{lccc}
    \toprule
    Model\hspace{4.5cm}   & 
    CER \\
    \midrule
    Ground Truth &  4.91 \\
    \midrule
    VITS with Phoneme Sequences &  4.26 \\
    VITS with Normalized Texts &  5.07 \\
    \midrule
    VITS2 with Phoneme Sequences &  \textbf{3.92} \\
    VITS2 with Normalized Texts &  \textbf{4.01} \\
    \bottomrule
  \end{tabular}
  \vspace{-0.5\baselineskip}
  \end{center}
\end{table}

\subsection{Comparison of Synthesis and Training Speed}
\label{ssec:results_synspd}
We compared our model's synthesis and training speed with those of the previous work~\cite{kim2021conditional}. We measured the synchronized elapsed time over the entire process to generate raw waveforms from input sequences with 500 sentences randomly selected from the LJ Speech dataset. We used a single NVIDIA V100 GPU with a batch size of 1. We also measured and averaged the elapsed time for the training computation of each step for five epochs on four NVIDIA V100 GPUs. Table~\ref{tab:sspeed} shows the results. 
As the duration predictor is more efficient and can be trained separately and the input sequences are shorter than in the previous work, its training and synthesis speed are improved; the improvements are 20.5\% and 22.7\%, respectively.

\newcommand{\mlcell}[2][p{2cm}c]{%
  \begin{tabular}[#1]{@{}c@{}}#2\end{tabular}}

\begin{table}[t]
  \caption{Comparison of the synthesis and training speed. The columns '$n$~kHz' and 'Real-time' of 'Synthesis' column denote the model's ability to generate $n\times1000$ raw audio samples per second and the synthesis speed over real-time, respectively. The 'sec/step' value of 'Training' column denotes the average elapsed time for the computation of training per step.}
  \label{tab:sspeed}
  \vspace{-0.5\baselineskip}
  \begin{center}
  \vskip -0.1in
  \begin{tabular}{lccc}
    \toprule
    Model\hspace{0.8cm}   & 
    \multicolumn{2}{c}{\mlcell{Synthesis\\ \hspace{0.18em} kHz \quad\quad Real-time}}\hspace{0.5em} & 
    \mlcell{Training\\sec/step}\hspace{0.1em} \\
    \midrule
    VITS & 1,779 & $\times$80.68  & 1.227  \\
    VITS2 & \textbf{2,144} & \textbf{$\times$97.25}  & \textbf{0.951}  \\
    \bottomrule
  \end{tabular}
  \vspace{-0.5\baselineskip}
  \end{center}
  \vskip -0.15in
\end{table}

\section{Conclusion}
\label{sec:Conclusion}
We propose VITS2, a single-stage text-to-speech model that can efficiently synthesize more natural speech. We improved the training and inference efficiency and naturalness by introducing adversarial learning into the duration predictor.
The transformer block was added to the normalizing flows to capture the long-term dependency when transforming the distribution.
The synthesis quality was improved by incorporating Gaussian noise into the alignment search.
The dependency on phoneme conversion, which was posing a challenge in achieving a fully end-to-end single-stage speech synthesis, was significantly reduced. The test results also show that overall intelligibility was improved. 
We demonstrated the validity of our proposed methods through experiments, quality evaluation, and computation speed measurement. Various problems still exist in the field of speech synthesis that must be addressed, and we hope that our work can be a basis for future research.
\section{Acknowledgements}
\label{sec:Acknowledgements}

We would like to express our gratitude to Jaehyeon Kim, Juhee Son, and Jungsu Hwang for their insightful discussions and valuable advice.



\bibliographystyle{IEEEtran}
\bibliography{mybib}

\end{document}